\documentclass[11pt]{elsarticle}
\usepackage{graphicx}
\usepackage{dcolumn}
\usepackage{bm}
\usepackage{amssymb}
\usepackage[pdftex]{color}
\usepackage{float}
\usepackage{lineno,hyperref}
\usepackage{epsfig}
\usepackage{booktabs} 
\usepackage{array} 
\usepackage{paralist}
\usepackage[english]{babel}
\usepackage{tikz}        
\usetikzlibrary{decorations.pathmorphing}
\usepackage{xcolor}
\usepackage{pgfplots}
\usepackage[export]{adjustbox}

\usepackage{geometry} 
\usepackage{amsmath}

\usepackage{subfig} 
\usepackage{fancyhdr} 
\usepackage[titles,subfigure]{tocloft} 

\newcommand{\0} {\textbf{0}}

\newcommand{\dvg} {\texttt{div}\,}

\newcommand {\Bc}  {\mathcal{B}}
\newcommand {\Cc}  {\mathcal{C}}

\newcommand {\Gc}  {\mathcal{G}}
\newcommand {\Hc}  {\mathcal{H}}

\newcommand {\Jc}  {\mathcal{J}}

\newcommand {\Pc}  {\mathcal{P}}
\newcommand {\Qc}  {Q_0}
\newcommand {\Rc}  {\mathcal{R}}

\newcommand {\Tc}  {\mathcal{T}}
\newcommand {\Uc}  {\mathcal{U}}
\newcommand {\Vc}  {\mathcal{V}}

\newcommand {\eb} {\mathbf{e}}
\newcommand {\fb} {\mathbf{f}}

\newcommand {\hb} {\mathbf{h}}

\newcommand {\mb} {\mathbf{m}}
\newcommand {\nb} {\mathbf{n}}

\newcommand {\qb} {\mathbf{q}}

\newcommand {\ub} {\mathbf{u}}

\newcommand {\xb} {\mathbf{x}}
\newcommand {\yb} {\mathbf{y}}

\newcommand {\Ab} {\mathbf{A}}

\newcommand {\Cb} {\mathbf{C}}

\newcommand {\Fb} {\mathbf{F}}

\newcommand {\Ib} {\mathbf{I}}
\newcommand {\Mb} {\mathbf{M}}

\newcommand {\Sb} {\mathbf{S}}

\newcommand {\Xb} {\mathbf{X}}

\def\gi{{\Gc_0}}
\def\gip{{\Gc'_0}}
\def\pt{\partial}

\definecolor{green}{RGB}{0,128,0}

\newcommand{\Th}{\Theta}

\journal{Journal of the Mechanics and Physics of Solids}
\begin{document}

\begin{frontmatter}

%
\title{Dehydration induced mechanical instabilities in active elastic spherical shells}
\author[r1]{Michele Curatolo}
\ead{paola.nardinocchi@uniroma1.it}
\author[r2]{Gaetano Napoli}
\ead{gaetano.napoli@unisalento.it}
\author[r1]{Paola Nardinocchi\corref{cor1}}
\ead{paola.nardinocchi@uniroma1.it}
\author[r3]{Stefano Turzi}
\ead{stefano.turzi@polimi.it}

\cortext[cor1]{Corresponding Author: 
P. Nardinocchi, Dipartimento di Ingegneria Strutturale e Geotecnica, via Eudossiana 18, I-00184 Roma, tel: 0039 06 44585242, fax: 0039 06 4884852.
}
\address[r1]{Sapienza Universit\`a di Roma, I-00184 Roma, Italy}
\address[r2]{Dipartimento di Matematica e Fisica ``E. De Giorgi'', Universit\`a del Salento, Lecce, Italy}
\address[r3]{Dipartimento di Matematica, Politecnico di Milano, Milano, Italy}
\begin{abstract}
Active-elastic instabilities are common phenomena in the natural world which have the aspect of sudden mechanical morphings. Frequently, the driving force of the instability mechanisms has a chemo-mechanical nature which makes these kind of instabilities very different from standard elastic instabilities. In this paper, we  describe and study the active-elastic instability occurring  in a  swollen spherical closed shell, bounding a water filled cavity, during a de-hydration process. The description is given through the outcomes of a few numerical experiments based on a stress-diffusion model which allows to glance at the phenomenon. The study is carried on  from a chemo-mechanical perspective through a few simplifying assumptions which allow to derive a semi-analytical model which takes into account both the stress state and the water concentration into the walls of the shell at the onset of the instability. Moreover, also the invariance of the cavity volume at the onset of instability, which is due to the impossibility to instantaneously change the cavity volume filled with water, is considered. It is shown as  the semi-analytic model matches very well the outcome of  the numerical experiments.
It is also shown as a wider range of mechanical instabilities can be produced when inhomogeneous shells are considered, even when the inhomogeneities can be described by a small number of parameters.
\end{abstract}


\end{frontmatter}

%
\section{\label{sec:1} Introduction}
%
%
Soft capsules confining microscopic cavities are common in Nature. Cavities can be water-filled, as is the case of the Fern Sporangium \cite{Noblin:2012,Llorens:2016}, or not, as is the case of Spagnum Moss \cite{Whitaker:2010}, just to cite a few. In both the cases, capsules undergo a dehydration process which determines the conditions to produce spores dispersion. The working principles of these mechanisms in Nature have been classified and studied in terms of the specific functional demands that these mechanisms fulfill. On the contrary, analyzing the possibility to reproduce them in soft polymers, which requires  an accurate modeling   and the identification of the determinants of the key mechanism, is still lacking.\\
Inspired by these observations, we investigate dehydration processes in spherical gel capsules going from a fully wet state with the cavities filled with water towards a dry state when exposed to air. The analysis started from numerical experiments based on a multi-physics three-dimensional model of stress-diffusion which showed the onset of mechanical instabilities during the dehydration process \cite{JMPS:2013,SM:2014,JAP:2014, Curatolo:2018, Curatolo:2018_pn}.\\
Mechanical instabilities in polymer gels have been extensively studied in the recent years with reference to swelling-induced surface instability of confined hydrogel layers on substrates \cite{Mora:2006,Kang:2010,Li:2013,Weiss:2013} and to  transient instabilities occurring during swelling processes \cite{Pandey:2013,SM:2014,Toh:2015,Takahashi:2016,Bertrand:2016,Curatolo:2019}.  The phenomenon we aim to describe is different and resembles the classical mechanical instabilities of pressurized spherical shells which has been largely investigating since $'50$ \cite{Green:1952, Wesolowski:1967, Haughton:1978, Hutchinson:2016}, and has been recently  having a new exciting life \cite{Pezzulla:2018,Holmes:2020}. However, our problem  presents a few characteristics which make it  distinguishable from the classical ones.\\
Firstly, in our problem the  external pressure, which is the control parameter in the classical stability analyses, is low and insignificant. On the contrary,  dehydration processes make spherical shells subject to a negative inner pressure, called \textit{suction pressure}, which is an unknown of the stress-diffusion problem, changes in time and can be considered as a live more than a dead load. Hence,  load conditions  are quite different from those considered in the Literature.\\  
Secondly, the driving force of the instability is the drying process which is controlled by the chemical potential of the environment, that is, the control of the process is not the mechanical pressure. Similar conditions have been studied in  \cite{Pezzulla:2018}, where the effect of spontaneous curvature, driven by differential growth, on the instability of spherical shells  has been investigated, within the context of non-euclidean theory of shells, through  a rational approach which allows to reduce the spontaneous curvature to an effective pressure-like dead load. 
In \cite{Pezzulla:2018}, it has also been shown as  a positive curvature corresponds to a positive external pressure (or, equivalently, to a negative inner pressure) causing a compression of the shell and possibly also a change in the cavity volume.
For dehydrating spherical shells, as those studied in the present paper, a spontaneous curvature  may be identified in terms of the change of dehydration degree across the thickness of the shell and it will result in a positive curvature for outer layers less hydrated  than inner layers, as the dehydration process starts from the outer layers.  Fundamentally different residual stress states drive  pressure and curvature buckling. The same is true when dehydration processes hold and the cavity is filled with liquid, that is, the cavity volume is constrained and it is expected that the buckling strategy of the shell is affected by the impossibility to change the cavity volume. Within the limits of the non euclidean shell theory, taking into account this further condition may be interesting and represents one of the future directions we identified in the Conclusion of the present paper.\\
The buckling of elastic spherical shells under osmotic pressure with the osmolyte concentration of the exterior solution as a control parameter has been studied in \cite{Knoche:2014}. Therein, the Authors  present a quantitative model aimed to capture the influence of shell elasticity on the onset of instability. Interestingly, they apply their model under cavity volume control assuming that the capsule volume can be considered as fixed when it is filled with an incompressible liquid that can leave the cavity on a very slow time scale like in drying mechanism.
That is the characteristic of our problem where the instability  occurs instantaneously with respect to the times of the diffusion which only can induce a change in the liquid content of the cavity. However, our model goes beyond as we present an instability study which holds for thin and thick shells, is based on the incremental analysis of both the mechanical and chemical equations which rule stress-diffusion in polymer gels, also including the cavity volume constraint.\\ 
In particular, after the description of  the de-hydration process which affects a closed spherical shell in terms of a three-dimensional stress-diffusion model, we evidence and numerically investigate  the onset of mechanical instabilities which are driven by the de-hydration process. As already evidenced,  even if instabilities occur when a critical pressure is attained, that pressure is not a control parameter of the instability process  which is driven by the de-hydration process which, on its turn, is controlled by the chemical potential of the external environment, the actual control parameter of the de-hydration and instability process. It motivates our choice to  analyse the instability problem  from a chemo-mechanical perspective through a few simplifying assumptions which allow to derive a semi-analytical model which takes into account both the stress state and the water concentration into the walls of the shell at the onset of the instability. Specifically, the stability analysis which we propose is borrowed from the study of elastic thick-walled spherical shells loaded by external pressure presented in \cite{Wesolowski:1967}. However, we extended that analysis to include the effects of water diffusion across the walls of the shell and the invariance of the cavity volume at the onset of instability. We show how the semi-analytic model matches the outcome of  the numerical experiments, based on the implementation of the stress-diffusion model,  and allows to  have a fast glance at mechanical instabilities of the shells numerically investigated and discussed in Section \ref{S3}.
We also show as a wider range of mechanical instabilities can be produced when inhomogeneous shells are considered, even when the inhomogeneities can be described by a small number of parameters.   
\section{Chemo-mechanical states of gels}
\label{S2}
%
The analysis of de-hydration processes starts from swollen gel bodies, however, it is convenient to introduce the dry state $\Bc_d$ of such bodies, use it as reference state  and describe the  chemo-mechanical state of gel bodies by a displacement field $\ub_d$ from the dry state and a water concentration $c_d$ per unit dry volume.
The displacement $\ub_d$ gives the actual position $X_d+\ub_d(X_d,t)$  of a point $X_d\in\Bc_d$, at time $t$,  whereas the water concentration $c_d$ gives the moles of water per unit dry volume at $X_d+\ub_d(X_d,t)$.\\
We assume that the free energy $\psi$ per unit dry-volume  depends  on the deformation gradient $\Fb_d=\Ib+\nabla\ub_d$ from $\Bc_d$  through an elastic component $\psi_e$, and on  water concentration $c_d$ per unit dry volume through a polymer--water mixing energy $\psi_m$: $\psi=\psi_e+\psi_m$, as prescribed by the Flory--Rehner thermodynamic model \cite{FR1,FR2}. As usual, we assume that  
any change in volume of the gel is accompanied by an equivalent uptake or release of water content, that is, 
\begin{equation}\label{Vconstraint}
J_d = \det\Fb_d = \hat J_d(c_d)=1+\Omega c_d\,,
\end{equation}
with $\Omega$  the molar volume of the water. Equation \eqref{Vconstraint} introduces a coupling between the state variables of the problem and is usually known as \textit{volumetric} or \textit{incompressibility} constraint. The volumetric constraint contributes to the definition of a relaxed free--energy $\psi_r$ as:
\begin{equation}\label{isoFR}
\psi_r(\Fb_d,c_d,p) = \frac{G_d}{2}(\Fb_d\cdot\Fb_d-3) + \frac{{\Rc} T}{\Omega}\,h(c_d) -p(J_d-\hat J_d(c_d))\,,
\end{equation}
with 
\begin{equation}\label{FRm2}
h(c_d)=\Omega\,c_d\,\textrm{log}\frac{\Omega\,c_d}{1+\Omega\,c_d} + \chi\,\frac{\Omega\,c_d}{1+\Omega\,c_d}\,,
\end{equation}
and the pressure $p$ as the reaction to the volumetric constraint,
which maintains the volume change $J_d$ due to the displacement equal to $\hat J_d(c_d)$ due to solvent absorption or release. Standard thermodynamical issues allow to derive the  constitutive equations for the dry-reference stress $\Sb_d$ (J/m$^3$) (the stress at the dry configuration $\Bc_d$)
and for the chemical potential $\mu$ (J/mol):
\begin{eqnarray}
\Sb_d&=&\hat\Sb_d(\Fb_d)-p\,\Fb_d^\star = G_d\,\Fb_d-p\,\Fb_d^\star\,,\quad \Fb_d^\star = J_d\,\Fb_d^{-T}\,,\nonumber\\
\mu&=&\hat\mu(c_d)+p\,\Omega = \Rc\,T\Bigl(\textrm{log}\frac{J_d-1}{J_d} + \frac{1}{J_d} + \frac{\chi}{J_d^2}\Bigr)+p\,\Omega\,,\label{csteqn}
\label{mud}
\end{eqnarray}
where, with a light abuse of notation, we wrote the relation for the chemical potential $\mu=\hat\mu(c_d)$   as $\hat\mu(J_d)$, by exploiting the volumetric constraint (\ref{Vconstraint}). Both stress and chemical potential consist of a constitutively determined component and a reactive component which couples the two main dynamical subjects of the theory. The components $\hat\mu(c_d)$ and $p\,\Omega$ are the mixing and mechanical contribution to  the chemical potential; the first is usually called \textit{osmotic pressure}.\\  
With these choices, the dissipation principle is reduced to the following inequality:
\begin{equation}\label{hmu}
\hb_d(\Fb_d,c_d,p)\cdot\nabla\mu(c_d,p)\le 0\,,\quad\mu(c_d,p)=\hat\mu(c_d)+p\,\Omega\,,
\end{equation}
being $\hb_d$ (mol/(m$^2$ s)) the  reference solvent flux, and is satisfied by assuming that
\begin{equation}\label{fh}
\hb_d = \hb_d(\Fb_d,c_d,p) = -\Mb(\Fb_d,c_d)\nabla(\hat\mu(c_d)+p\,\Omega)\,,
\end{equation}
with
the diffusion tensor $\Mb(\Fb_d,c_d) $(mol$^2$/(s m J)) as a positive definite tensor. In particular, we also assume that $\Mb$ is isotropic and linearly dependent on $c_d$, and diffusion always remains isotropic during any process \cite{Hong:2008,Zhang:2009,Chester:2010,JMPS:2013}. These assumptions determine the representation of the diffusion tensor in terms of the inverse of the Cauchy--Green strain tensor $\Cb_d=\Fb_d^T\Fb_d$ as
\begin{equation}\label{Mh}
\Mb(\Fb_d,c_d) =  \frac{D}{{\Rc}T}\,c_d\Cb_d^{-1}\,,
\end{equation}
with $D$ (m$^2$/s) the diffusivity. 
Finally, the balance equations of the model are:
\begin{equation}\label{bal1}
0=\dvg\,\Sb_d \quad\textrm{and}\quad
\dot c_d = -\dvg\,\hb_d \,,
\end{equation}
 on $\Bc_d\times\Tc$. They are supplemented by the boundary conditions on  $\partial_t\Bc_d\times\Tc$ and $\partial_u\Bc_d\times\Tc$
\begin{equation}\label{bal2}
   \Sb_d\,\mb=-\bar p\Fb_d^\star\mb\quad\textrm{and}\quad
    \ub_d=\bar\ub\,, 
\end{equation}
respectively, where we only considered boundary pressure $\bar p$ and set $\bar\ub$ for the assigned displacement; and on $\partial_q\Bc_d\times\Tc$ and $\partial_c\Bc_d\times\Tc$
\begin{equation}\label{bal3}
   -\hb_d\cdot\mb = q_s\quad\textrm{and}\quad
   \hat\mu(c_s)+p\,\Omega=\mu_e\,,
\end{equation}
respectively, with $q_s$ the boundary flux and $c_s$ the concentration  field on $\partial_c\Bc_d$ which is assigned implicitly by controlling the external chemical potential $\mu_e$. The initial conditions 
\begin{equation}\label{bal4}
\ub_d = \ub_{do}\quad\textrm{and}\quad c_d=c_{do} \,,
\end{equation}
on $\Bc_d\times\{0\}$ make the problem doable: $\ub_{do}$ and $c_{do}$ are the initial values of the fields $\ub_d$ and $c_d$, respectively. Everywhere,  a dot denotes the time derivative and $\texttt{div}$ the divergence operator.
\subsection{De-hydration of gel capsules}
We discuss de-hydration of spherical shells confining spherical cavities.
\begin{figure}[h]
\centering\includegraphics[width=5in]{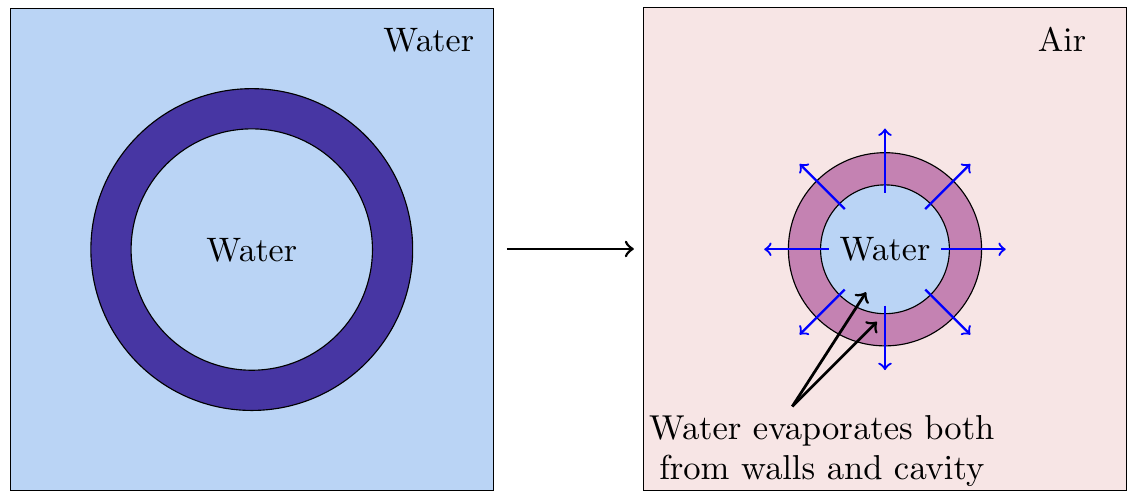}
\caption{A sketch of the de-hydration process. Initial steady stress-free swollen state of the spherical capsule: water fills the cavity and the external environment (left). After exposition to air, the de-hydration process starts and  water in the walls and in the cavity moves towards the outer environment (right).}
\label{fig:1} 
\end{figure}
The dry system $\Bc_d$  is a spherical shell of external radius $R_d$ and thickness $H_d=R_d-R_c$, with $R_c$  the radius of the cavity $\Cc_d$. The shell size increases to accomodate an amount of water which is determined by the shear modulus $G_d$ and the Flory parameter $\chi$ through the equation
\begin{equation}\label{steady}
\Rc\,T\Bigl(\textrm{log}\frac{\lambda_o^3-1}{\lambda_o^3} + \frac{1}{\lambda_o^3} + \frac{\chi}{\lambda_o^6}\Bigr) + \frac{G}{\lambda_o}\Omega =\mu_{o}\,,
\end{equation}
corresponding to the equilibrium conditions $\Sb_d=\0$ and $\mu=\mu_e$ with $\mu_e=\mu_o$.\footnote{With these, the balance equations \eqref{bal1} are trivially satisfied.} We denote this steady and stress-free swollen state as $\Bc_o$: the shell has radius  $\lambda_o\,R_d$ whereas the cavity $\Cc_o$, assumed to be completely filled with water, has radius $\lambda_o\,R_c$.\\
We assume that this state represents the initial state of the system under a de-hydration process which starts from the fully swollen state $\Bc_o$ and proceeds by de-hydrating the body from the outside. It corresponds to pull out the swollen spherical shells, with its cavity filled with water, from the bath and to expose it to air  (see cartoon  in figure \ref{fig:1}). Diffusion starts and water is expelled from both the gel and the cavity. Being water incompressible, the cavity volume must always be equal to the volume of the water it contains;
thus, when water is pumped out of the cavity, the cavity volume reduces and the cavity wall $\partial_i\Bc_d=\partial\Cc_d$
may be pulled by an increasing negative pressure.\\
From the modelling point of view, exposing the capsule to air means changing the chemical potential 
at the external boundary $\partial_e\Bc_d$ from $\mu_o$ to  $\mu_e<\mu_o$. If it is the case,  equations \eqref{bal1}-\eqref{bal4} allow to follow the dynamics of the process.
We assume that all along the process the cavity stays always  filled with solvent\footnote{It corresponds to assume that no delamination of liquid from cavity walls can occur, or equivalently, that surface energy per unit area of the cavity is much higher than stretching modulus $G\lambda_o(R_d-R_c)$.} and the chemical potential on $\partial\Cc_d$ is determined by $\mu_i=\Omega\,p_i(t)$ with  the pressure term $p_i$, representing the suction pressure. On the other side, we assume that the outer environment is filled with air, that is,  an ideal gas whose content in water determines the value of the chemical potential which can be related to the relative humidity of the air, and set  $\mu_e=\hat\mu_e(t)+p_e$ with $\hat\mu_e(t)$ the control law of the problem and the base atmospheric pressure $p_e=0$. So, in the end, we write down:
\begin{equation}
 \mu_e=\hat\mu_e(t)\quad\textrm{on}\quad\partial_e\Bc_d\quad
 \textrm{and}\quad
 \mu_i=\Omega\,p_i(t)\quad\textrm{on}\quad\partial\Cc_d\,,
 \label{BCc}
\end{equation}
and
\begin{equation}\label{BCt}
   \Sb_d\,\mb=- p_e\,\Fb_d^\star\,\mb=\0\quad\textrm{on}\quad\partial_e\Bc_d\times\Tc\quad\textrm{and}\quad 
    \Sb_d\,\mb=- p_i\,\Fb_d^\star\,\mb\quad\textrm{on}\quad\partial\Cc_d\times\Tc\,,
\end{equation}
where $\Fb_d^\star= J_d \Fb_d^{-T}$ denotes the adjugate of the deformation gradient.\\
The suction pressure $p_i=p_i(t)$, a key ingredient in the onset of instabilities, is modelled as the reaction to the volumetric coupling which relates
the volume $v_s^c=v_s^c(t)$ of the solvent in the cavity to the volume of the cavity $v_c=v_c(t)$, which can simply be written down as: at each instant $\,t\in\Tc$ as solvent flows out of the cavity, it must hold
\begin{equation}\label{vol_cavity}
v_c(t)=v_s^c(t)\,.  
\end{equation}
It is worth noting that the global constraint (\ref{vol_cavity}) adds a further coupling between the state variables of the multiphysics problem other than the common local volumetric constraint (\ref{Vconstraint}).
Constraint (\ref{vol_cavity}) can be enforced by considering the augmented total free-energy defined by
\begin{equation}\label{inner_p}
\int_{\Bc_d} \psi_r\,dV_d - p_i\,(v_c - v_s^c)\,,
\end{equation}
so that the cavity pressure $p_i$ can be viewed as the Lagrange multiplier enforcing the constraint.
The cavity volume $v_c$ depends on the actual configuration $\Cc_t$ of the cavity
at time $t$, and can be measured by evaluating the following integral
\begin{equation}\label{vi}
v_c(t)=\int_{\Cc_t}dv =-\frac{1}{3}\int_{\partial_i\Bc(t)} x\cdot\nb\, da
                   =-\frac{1}{3}\int_{\partial_i\Bc(t)}(X_d+\ub_d)\cdot\Fb_d^\star\,\mb\, dA_d\,,
\end{equation}
with $\nb$ the normal to the actual boundary $\partial_i\Bc(t)=f(\partial\Cc_d)$.\footnote{We note that the internal boundary of the gel 
$\partial_i\Bc_d$ coincides with the boundary of the cavity $\partial\Cc_d$, 
proviso an opposite orientation of the normal.}
The water volume at time $t$ is the sum of the initial water content $v_s^c(0)$
of the cavity, plus the water volume $Q_i(t)$ that crossed the cavity boundary during the 
time interval $(0,t)$, that is, $v_s^c(t)=v_s^c(0)+Q_i(t)$.
The initial water content equals the initial cavity volume $v_{co}=v_c(0)$, that is,  from  (\ref{vi}), 
it holds
\begin{equation}
v_s^c(0)=v_c(0)=-\frac{1}{3}\int_{\partial_i\Bc_d}(X_d+\ub_o)\cdot\Fb_o^\star\,\mb\, dA_d\,,
\end{equation}
with $\Fb_o^\star= J_o\,\Fb_o^{-T}$ and $J_o$  the adjugate and the Jacobian determinant of the initial swollen deformation gradient $\Fb_o=\lambda_o\Ib$. The water volume $Q_i(t)$ that  crossed the cavity boundary  and was absorbed by the gel can be evaluated by:
\begin{equation}\label{qi}
Q_i(t)=\int_0^t\, \dot Q_i(\tau)\,d\tau
      =\Omega\int_0^t\,\left(\, \int_{\partial_i\Bc_d}q\, dA_d\, \right)\,d\tau
      = -\Omega\int_0^t\,\left(\, \int_{\partial_i\Bc_d}\hb_d\cdot\mb\, dA_d\, \right)\,d\tau\,.
\end{equation}
Equations \eqref{vi}-\eqref{qi} allow to follow the de-hydration process of the spherical capsule.
It is worth noting that the effects of the process on the mechanics of the shell are very different depending on the shear modulus of the polymer, once fixed the liquid-polymer affinity. Indeed, high or low shear moduli $G_d$  identify the initial state $\Bc_o$ as a poorly or highly swollen state and can determine a very different dynamics\cite{Curatolo:2018,Curatolo:2020}.  For highly swollen gels, due to the great amount of liquid inside shell walls, the de-hydration process starts with the liquid  firstly released from the shell rather than from the cavity. As a consequence, suction effect does not become immediately apparent and the inner pressure $p_i$ takes non-negative values. By contrast, for poorly swollen gels, liquid is mainly released from the cavity and the inner pressure quickly attains  negative values \cite{Curatolo:2020}, a condition which is determinant for the onset of mechanical instabilities, as we'll discuss in the rest of the paper.
\section{A glance at active elastic instabilities}\label{Instabilities}
\label{S3}
The effects of the de-hydration process on shell shape is numerically studied; the onset of the so-called \textit{active elastic} instabilities is investigated and the phenomenology is  contrasted with previous results which some of the Authors got for cubic capsules \cite{Curatolo:2018}.\\
In the following numerical experiments, we fixed the set of material parameters listed in Table \ref{tab:1}. With these choices, the dimensionless parameter $\epsilon=G\Omega/RT$, which is the ratio between the two key material constants of the mechanical and chemical free-energy is around $0.37$. The value of $\varepsilon\gtrsim 10^{-1}$ and of the affinity parameter $\chi\lesssim 0.8$ allows to infer that the gel is poorly swollen.
%
\begin{table}[h]
\centering
\begin{tabular}{ll}
\hline
Shear modulus	& $G_d = 5\cdot 10^7$ Pa; \\[1mm]
\hline
Flory parameter	& $\chi  = 0.2$; \\[1mm]
\hline
Water molar volume	& $\Omega =1.8\times 10^{-5}$ m$^3$/mol;\\[1mm]
\hline
Water diffusivity     & $D  = 10^{-9} $m$^2$/s; \\[1mm]
\hline
Temperature     & $T  = 293 $K; \\[1mm]
\hline
\end{tabular}
\caption{\label{tab:1} Values of parameters used in numerical experiments.}
\end{table} 
Indeed, fixed $\mu_o=0$ $[$J$/$mol$]$, equation \eqref{steady} yields the value $\lambda_o= 1.152$ of the swelling ratio which corresponds to a $15\%$  increase of the capsule thickness and radius which change from the dry values $H_d=1.25\cdot 10^{-3}$ m and $R_d=1\cdot 10^{-2}$ m to the swollen values $H_o=1.44\cdot 10^{-3}$ m and $R_o=1.152\cdot 10^{-2}$ m. The corresponding initial values for the displacement and pressure field are  $\ub_o=(\lambda_o-1)\Xb_d$ and $p_o=4.338\cdot 10^7$ Pa. From the value $\mu_o$, the chemical potential is made to change following a time law
$\hat\mu_e(t)$ which brings the value of the external chemical potential from the initial value $\mu_0$ to the final value $\mu_f$ in $t_\mu=2500$ through a  smoothed step function; then, the final value $\mu_f$ is kept fixed.\\ 
\begin{figure}[h]
\centering\includegraphics[width=5.5in]{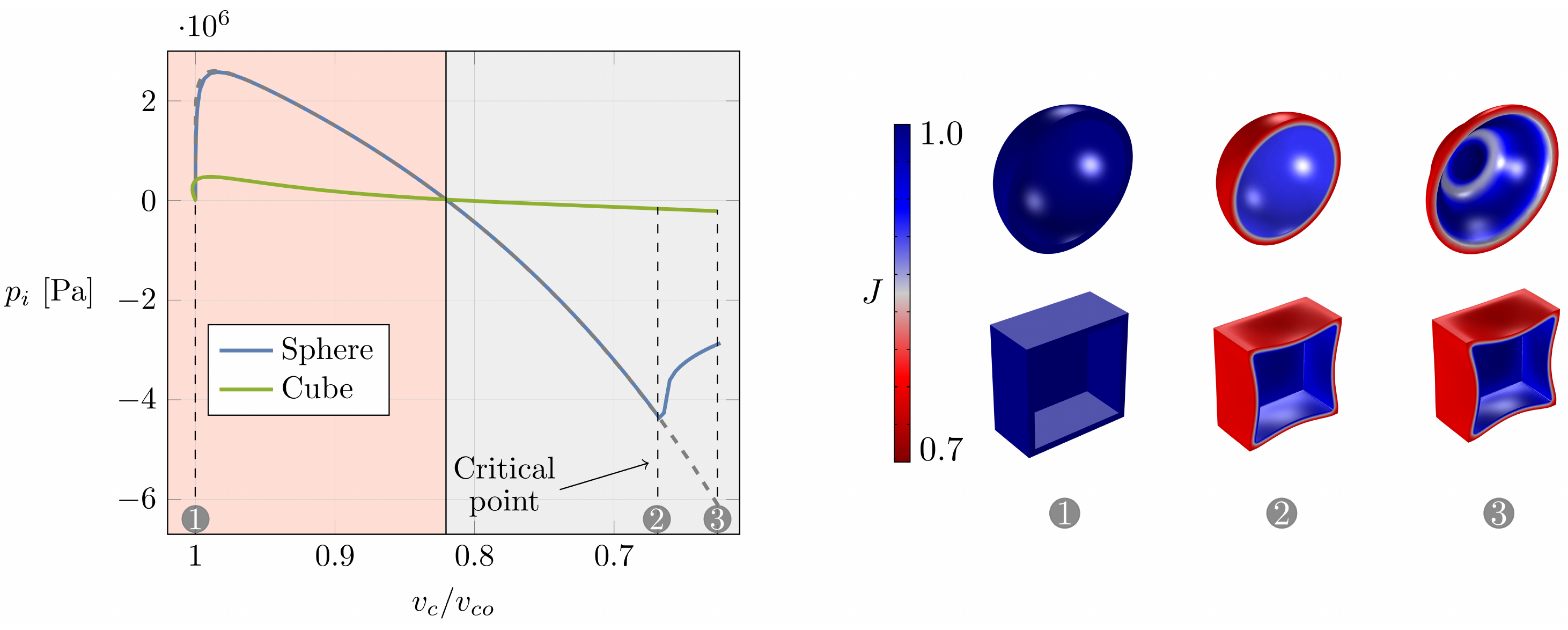}
\caption{Pressure--volume curves for spherical (blue) and cubic (green) gel capsules: pink (grey) background identifies the first phase of the process when inner pressure takes positive (negative) values (left). Cubic and spherical  capsules are shown at different values of $v_c/v_{co}=1, 0.68 , 0.65$ and allow to appreciate the change in shape occurred in the capsules.}
\label{fig:2}
\end{figure}
The pressure-volume curves shown in figure \ref{fig:2} (left) allow to highlight the observed dynamics in spherical capsules (blue line) and to evidence the onset of a mechanical instability which is, on the contrary, not observed in cubic capsules of similar size (green line).  At the beginning of the de-hydration process from the swollen shell, liquid is mainly expelled from the cavity walls and pressure changes at almost unchanged cavity volume, as the initial deep slope of the blue line in figure \ref{fig:2} (left) shows. As diffusion goes on, water is released from both the shells and the cavity and the cavity volume reduces; we follow it until a decrease of about $30\%$ is attained, corresponding to $v_c/v_{co}=0.7$. In a first phase (pink background), inner pressure takes positive values  which, at the same dimensionless cavity volume ratio $v_c/v_{co}$, are higher for spherical than for cubic capsules, and breathing modes can be observed in both the situations  \cite{Curatolo:2018}. In a second phase (grey background), inner pressure takes negative values, so realizing the so-called suction effects on the walls of the capsules. Whereas the walls of cubic capsules bend under negative pressure,  spherical capsules, made stiffer by the geometrical symmetry, do not bend, as it is shown in figure \ref{fig:2} (right) by the cartoons corresponding to number $2$. Moreover, as it is energetically very expensive reducing cavity volume in spherical capsules, we also observe higher values of the negative pressure at the same value of the ratio $v_c/v_{co}$.\\
At a critical value of the inner pressure, a mechanical instability is observed which   allows to release the elastic energy accumulated in the shell during the process. The onset of the mechanical instability changes the shape of the sphere very sharply and the pressure-volume curve shows an almost vertical slope  at the critical point, as it is evidenced in figure  \ref{fig:2} (left). Indeed, as the cavity is still filled with water, and diffusion is typically a slow process (here, the characteristic diffusion time $\tau_d=H_o^2/D\varepsilon\simeq 10^3$ s), instability occurs at almost constant volume. Figure \ref{fig:2} (right) also shows spherical and cubic capsules at different values of $v_c/v_{co}=1, 0.68 , 0.65$, corresponding to the points $1, 2, 3$ evidenced in the pressure-volume diagram. At the value $v_c/v_{co}=0.65$ (point $3$), the cubic capsule shows an evident increased bending of its walls whereas the spherical capsule has attained a \textit{sombrero} shape. Key determinants of the mechanical instability  are the cavity volume ratio $v_c/v_{co}$ and the inner pressure $p_i$. In particular, we observe that the onset of instability corresponds to a pair  $v_c/v_{co}=0.68$ and $p_i=-4.2\cdot 10^6$ Pa. These values will be used as benchmark values in the following section where a semi-analytical study of the instability is presented.\\
Finally, the mechanical instability also affects water fluxes, an aspect of the release dynamics which can be noteworthy  in  applications requiring  a control on  fluxes  and is beyond the aim of the paper.
\section{Study of the chemo-mechanical instability}
\label{S4}
The key aspects of the instability problem which affects spherical shells during de-hydration processes can be described from a mechanics perspective through a few simplifying assumptions which allow to derive a semi-analytical model. The stability analysis which we propose in the following is borrowed from the study of elastic thick-walled spherical shells loaded by external pressure presented in \cite{Wesolowski:1967}, and is extended to consider the diffusion equation \eqref{bal1}$_2$ and the invariance of the cavity volume at the onset of instability.\\

\subsection{Spherical solution}
Before instability occurs, the shell is spherical and the chemo-mechanical state variables are determined as stationary solutions of equations \eqref{bal1}. Assuming the dry configuration as  the reference configuration, we consider purely radial deformations of the thick shell which are represented  as 
\begin{equation}
r=r(R)\,,\quad\theta=\Theta\,,\quad\phi=\Phi\,,
\label{sphdef}
\end{equation}
where $(R,\Theta, \Phi)$ and $(r, \theta, \phi)$ are the spherical coordinates of a point in the reference and  current configuration, respectively. We denote all the chemo-mechanical variables corresponding to the spherical solution with the subscript `$0$'; so  $\Fb_0=\textrm{diag}(r',r/R,r/R)$ is the deformation gradient corresponding to the deformation \eqref{sphdef},  with a prime denoting differentiation with respect to the radial coordinate $R$. The equilibrium configurations have to satisfy  the volumetric constraint \eqref{Vconstraint}: $\det \Fb_0=J_0(R)$, with $J_0 (R) = 1 + \Omega c_0(R) $, where $c_0$  denotes the solvent concentration in the spherical solution. When we substitute \eqref{sphdef} into the constraint, we get
\begin{equation} \label{eq:constraint}
r^2r'=R^2 J_0.
\end{equation}
Notice that, unlike the classical analysis \cite{Wesolowski:1967,BenAmar:2005}, where $J_0 =1$, here $J_0 $ is an unknown function of the radial coordinate. When $J_0 =1$, equation \eqref{eq:constraint} can be easily solved and yields the classical result $r(R) = (3 a_0 + R_c^3)^{1/3}$, where $a_0$ is an integration constant. By contrast, in our case, equation \eqref{eq:constraint} has to be solved together with the chemo-mechanical balance equations. 

%
 
Let us introduce  $\Qc(R):=R/r(R)$ so that  the deformation gradient of the spherical solution can be cast in the form
\begin{equation}
\Fb_0=\textrm{diag}(\Qc^2 J_0 ,\Qc^{-1},\Qc^{-1})\,.
\end{equation}
Then, according to the neo-Hookean hyperelastic model, the Piola-Kirchooff stress tensor can be written as 
\begin{equation} \label{eq:S0}
\Sb_0=\textrm{diag}(-p_0 \Qc^{-2} + G_d J_0  \Qc^{2} ,  - p_0 J_0  \Qc + G_d \Qc^{-1}, - p_0 J_0  \Qc + G_d \Qc^{-1})\,,
\end{equation}
where $p_0(R)$ is the Lagrangian multiplier related to the constraint $J_0  = 1 + \Omega c_0$. On the other hand, the representation formula of the chemical potential is unchanged by the spherical symmetry and it holds
\begin{equation} \label{eq:mu}
 \mu_0=\Rc\,T\Bigl(\textrm{log}\frac{J_0 -1}{J_0 } + \frac{1}{J_0 } + \frac{\chi}{J_0^2 }\Bigr)+ p_0\,\Omega\,.
\end{equation}
Once observed that $\Qc'=\Qc(1-J_0\Qc^3)R^{-1}$, the balance of forces $\dvg\Sb_0 =\0$ (in spherical coordinates, $R \partial_R S_{0_{RR}} + 2 S_{0_{RR}} -  S_{0_{\Th \Th}} -  S_{0_{\phi \phi }} = 0$) reduces to: 
\begin{equation}\label{bf}
 p_0' R + 2G_d\Qc(-1+J_0  \Qc^3)^2 - G_d \Qc^4 J_0 ' R =0\,.
\end{equation}
As far as the diffusion problem is concerning, we consider the quasi-static version of \eqref{bal1}$_2$, that is, $\dvg\hb_0 = 0$. Due to the spherical symmetry,  the solvent flux $\hb_0$ is purely radial, that is,  $ \hb_0 (R) = ( h_{0_R} (R), 0 , 0)$, and that balance equation, after a first integration, reduces to
\begin{equation}
R^2  h_{0{_R}} = C_0,
\label{eq:h0}
\end{equation}
where $C_0$ is an integration constant and the constitutive equations \eqref{fh}-\eqref{Mh}, reduced by the spherical symmetry, deliver 
\begin{equation} \label{eq:diff}
 h_{0{_R}}= - \frac{D}{\Qc^4 J_0 ^2}\left[\frac{-2 \chi (J_0  - 1) + J_0 }{\Omega J_0 ^3} J_0 ' + \frac{(J_0 -1)}{\Rc T} p_0' \right]\,.
\end{equation}
Finally, by using  the boundary conditions \eqref{BCc} and  \eqref{BCt}, we obtain the boundary conditions for the spherical problem:
\begin{subequations}
\begin{align}
& S_{0_{RR}}(R_d) =0 , \qquad  \mu(R_d) = \mu_e\,,  \label{eq:bc0e} \\
& \mu(R_c) + \Omega \Qc^2(R_c) S_{0_{RR}}(R_c) =0\,. \label{Pbeta}
\end{align}
\label{bc:0}
\end{subequations}
Equations \eqref{eq:bc0e} express the conditions of vanishing pressure and assigned chemical potential at the external boundary. Equation \eqref{Pbeta}
is derived from (\ref{BCc})$_2$ that relates the chemical potential $\mu(R_c)$ and the pressure $p_i$ on the inner boundary,  where the boundary condition \eqref{BCt}$_2$ has been used to express  $p_i$ in terms of the radial stress component as $p_i=-\Qc^2(R_c) S_{0_{RR}}(R_c)$.\\
Finally, it is worth noting that, at any time before instability occurs, the enclosed volume $v_c$  is determined by the liquid filling the cavity and the relationship between the radius $r_c$ of the cavity and the volume $v_c$,  is 
\begin{equation}
r_c=\left(\frac{3v_c}{4\pi}\right)^{1/3}\,.
\label{vcrc}
\end{equation}
Hence, whenever the enclosed volume is assigned, equation \eqref{vcrc} delivers  the boundary condition
\begin{equation}
r(R_c) = r_c.
\label{bc:r}
\end{equation}
\begin{figure}[h]
\centering{
\includegraphics[width=5.5in]{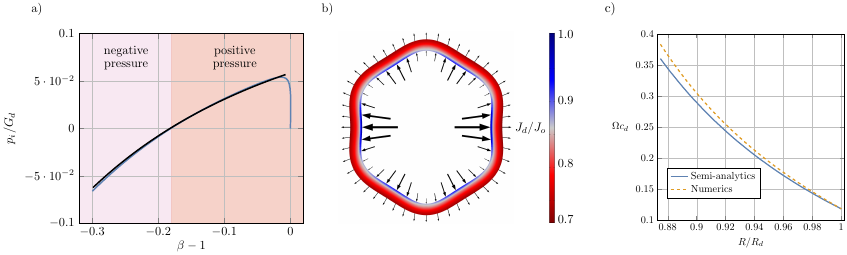}
}
\caption{Pressure--$\beta$ curves from numerics (green line) and analytics (black line). Shape of the shell at the critical point, with arrows denoting liquid flux (center). Water concentration across the shell thickness at the critical point (right). Light pink and pink portions evidence negative and positive pressure regimes.}
\label{fig:3}
\end{figure}

\noindent
If $\mu_e=0$ and the enclosed volume is not constrained, this spherical problem admits the stress-free {\it swollen solution} discussed in \S \ref{S2}. In fact, this is the only case in which $\mu(R)=0$ implies an uniform deformation $\lambda_o = \Qc^{-1}=J_0 ^{\frac{1}{3}}$, where $\lambda_o$ satisfies equation \eqref{steady}. The enclosed volume is then $v_{co} =  \frac{4}{3} \pi R^3_c J_0$.\\
Whenever the enclosed volume is assigned, the solution of the spherical problem determined by the equations \eqref{eq:constraint}, \eqref{bf}, \eqref{eq:h0}-\eqref{eq:diff}, with the boundary conditions \eqref{bc:0} and \eqref{bc:r}, provides the relationship between the inner pressure $p_i$ and $v_c$, which we write in the form $ p_i=p_i(\beta)$, with 
\begin{equation}
\beta := \frac{v_c}{v_{co}}
\label{eq:beta}
\end{equation}
the ratio between the current and the initial volume of the cavity. Interestingly, we find that $p_i$ has the same sign of $\beta - 1$; hence, as expected, for enclosed volumes smaller than $v_{co}$ (i.e., $\beta<1$), a negative pressure arises that produces a inner cavity compression. Therefore, the instability is expected for $\beta< 1$. 

Figure \ref{fig:3} (left) shows the dimensionless inner pressure $ p_i/G_d$ as a function of $\beta$ from analytics  (black line) and from numerics (green line), where the first consists in a quasi static analysis of the stress diffusion problem whereas the second also describes the transient behaviour. The two curves show a remarkable agreement where $\mu_e$ has reached its plateau value $\mu_e = - 2 \cdot 10^3 \, \text{J/mol}$. By contrast, in the transient region where $\mu_e$ suddently drops to its plateau value and the pressure rises at nearly constant cavity volume, as expected, the static solution cannot reproduce the numerical values. The shape of the shell at the critical point is shown in figure \ref{fig:3} (center). Arrows denote liquid flux and it can be appreciated how the instability affects liquid flux, an issue which is beyond the aim of the present paper. Finally, colour code scale represent the ratio $J_d/J_o$ measured in the numerical experiments. Figure \ref{fig:3} (right) shows the profile of water concentration across the shell thickness at the critical point and evidences as the quasi-static solution satisfactorily matches the numerical solution coming out from a dynamical analysis. 
\subsection{The linearized problem}
\label{sec:linearized}
In order to find the critical values of $\beta$ at which the instability occurs, we consider the axisymmetric incremental fields $u, v, p_1, J_1$ and write 
\begin{subequations}
\begin{align}
\xb(R,\Th) & =  r(R)\eb_R + \epsilon (u(R,\Th) \eb_R +  v(R,\Th) \eb_\Th)\, ,  \label{eq:uv1} \\
p (R,\Th) & =  p_0(R) + \epsilon p_1(R,\Th)\, ,   \\
J_d(R,\Th) & =  J_0(R) + \epsilon J_1(R,\Th)\, .  
\end{align}
\end{subequations}
Therefore, the incremental deformation gradient is
\begin{equation}
\Fb_1 = 
\begin{pmatrix}
\partial_R u \;\;\; & R^{-1} (\partial_R u - v) & 0\\ 
\partial_R v \;\;\; & R^{-1} (u + \partial_\Th v)& 0 \\
0 & 0 & R^{-1} (u +  v \cot \Th)
\end{pmatrix}\,,
\end{equation}
and the unknown fields $u$ and $v$ have to satisfy the volumetric constraint which at the first order is $J_0 \,\mathrm{tr}( \Fb_0^{-1} \Fb_1)= J_1$, and reads
\begin{equation}
 \Qc^{-2} \; \partial_R u + J_0 \Qc R^{-1} (2 u + \partial_\Th v + v \cot \Th) - J_1 = 0\,.
\label{inco}
\end{equation}
This equation have to be coupled with the incremental equilibrium equations
\begin{equation}
\dvg \; \Sb_1 =0, \qquad \dvg \; \hb_1 =0,
\label{eq:lin}
\end{equation}
where $\Sb_1$ and $\hb_1$ represent the linearized Piola-Kirchhoff stress tensor and solvent flux, respectively. According to the neo-Hookean model, the nonvanishing components of the incremental stress tensor $
\Sb_1 = - p_1 \Ib  -  p_0 \Fb_1^ {-T} + G_d  \Fb_1$ are:
\begin{subequations}
\begin{equation}
\Sb_{1_{RR}} = p_0 J_0^{-1} \Qc^{-4}(\pt_R u - Q^2 J_1) - p_1 \Qc^{-2} + G_d \pt_R u\,,
\end{equation}
\begin{equation}
\Sb_{1_{R\Th}} = p_0 \Qc^{-1} \pt_R v + G_d R^{-1} (\pt_R u - v)\,,
\end{equation}
\begin{equation}
\Sb_{1_{\Th R}} = p_0 \Qc^{-1}R^{-1} (\pt_R u - v) + G_d \pt_R v\,,
\end{equation}
\begin{equation}
\Sb_{1_{\Th \Th}} = - \Qc (p_1  J_0  +  p_0 J_1) + R^{-1}(G_d + J_0 p_0 \Qc^2)(u + \pt_\Th v)\,,
\end{equation}
\begin{equation}
\Sb_{1_{\Phi \Phi}} = - \Qc (p_1  J_0  +  p_0 J_1) + R^{-1}(G_d + J_0 p_0 \Qc^2)(u + v \cot \Th)\,.
\end{equation}
\label{S1}
\end{subequations}
Similarly, we consider the perturbation of the water flux up to the first order, $\hb_d(R,\Th) = \hb_0(R)+ \epsilon \hb_1 (R,\Th)$ and from  equation \eqref{fh}, we obtain $\hb_1 = - \Mb _0 \nabla \mu_1 - \Mb _1 \nabla \mu_0$ where 
\begin{align}
\Mb_0 & = \frac{D}{\Rc T} \frac{J_0 -1}{\Omega} \Fb_0^{-2}, \\
\Mb_1 & = \frac{D}{\Rc T \Omega}  \Fb_0^{-1}\left[J_1 \Ib - (J_0 -1) (\Fb_1 \Fb_0^{-1} + \Fb_0^{-1}\Fb_1^T) \right]\Fb_0^{-1}\,, \\
\mu_1 & = -\frac{\Rc T}{(J_0 - 1)J_0^3 } [2\chi(J_0 - 1) - J_0] J_1 + \Omega p_1\,,
\label{mu1}
 \end{align}
while $\mu_0(R)$ is given by eqn. \eqref{eq:mu}. Consequently, the nonvanishing components of $\hb_1$ are 
\begin{subequations}
\begin{equation}
h_{1_R} =  -\frac{D}{\Rc T \Omega J_0^3 \Qc^6}\left[\mu'_0   \Qc^2 J_0   J_1 +  (J_0  -1) (-2 \mu'_0  \pt_R u  +  J_0 \Qc^2 \pt_R \mu_1) \right]\,,
\end{equation}
\begin{equation}
h_{1_\Th} = \frac{D (J_0 -1)}{\Rc T \Omega J_0^2 \Qc^3 R}\left[-J_0^2 \Qc^5 \pt_\Th \mu_1 + \mu'_0 (J_0 \Qc^3 (\pt_\Th u  - v ) + R \pt_R v) \right]\,,
\end{equation}
\end{subequations}
and the incremental diffusion equation reduces to:
\begin{equation}
\frac{1}{R^2} \pt_R (R^2 h_{1_R}  )  + \frac{1}{R \sin \Th} \pt_\Th (h_{1_\Th} \sin \Th ) =0\,.
\label{eq:4.40}
\end{equation}
The nonvanishing components of the incremental equilibrium equations \eqref{eq:lin}  and the constraint equation \eqref{inco} provide a
system of $4$ coupled partial differential equations for $u$, $v$, $p_1$ and $J_1$ as a function of $R$ and 
$\Th$, where the coefficients depend on the finite-strain solution obtained at zeroth order.

To solve this problem, we expand the unknown fields in Legendre polynomials 
\begin{subequations}
\begin{align}
u(R,\Th) & = \sum_{l=1}^{\infty} \Uc_{l}(R) {\rm P}_l(\cos \Th)\,, 
& v(R,\Th)  & =  \sum_{l=1}^{\infty} \Vc_{l}(R)  {\partial_\Th [{\rm P}_l(\cos \Th)]}\,, \\
p_1(R,\Th) &= \sum_{l=1}^{\infty} \Pc_l(R) {\rm P}_l(\cos \Th)\,, 
& J_1(R,\Th) & = \sum_{l=1}^{\infty} \Jc_{l}(R)  {{\rm P}_l(\cos \Th)}\,.
\end{align}
\label{solu}
\end{subequations}
We do not consider the mode $l=0$ in the expansions since it corresponds to a symmetric increase in shell radius and its existence does not correspond to a true axisymmetric bifurcation. By separation of variables,  we obtain a systems of ordinary differential equations for $\Uc_l$, $\Vc_l$, $\Pc_l$ and $\Jc_l$.  This approach generalizes the classical ones for the stability of shells under pressure \cite{Wesolowski:1967} and of growing shells \cite{BenAmar:2005}.\\
%
%
We now use equation \eqref{inco} to obtain
\begin{equation}
\Jc_l =   R^{-1} J_0 \Qc [2 \Uc_l - l (l+1)\Vc_l] + \Qc^{-2}\Uc'_l,
\label{eqB}  
\end{equation}
and, therefore, eliminate $\Jc_l$ in the differential equations. Furthermore, to deal with \eqref{eq:4.40} it is convenient, from a computational standpoint, to consider the following expansion for $h_{1_{R}}$
\begin{equation}
\qquad h_{1_R}(R,\Th)  = \sum_{l=1}^{\infty} \Hc_l(R) {\rm P}_l(\cos \Th)\, ,
\label{soluh}
\end{equation}
and solve the system of coupled equations in terms of $\{\Uc_l, \Uc'_l, \Vc_l, \Vc'_l, \Pc_l, \Hc_l\}$. More precisely, we introduce the vector of unknowns $\qb _l = \{\Uc_l, \Uc'_l, \Vc_l, \Vc'_l, \Pc_l, \Hc_l\}$, so that our system of first order linear differential equations is cast into the form
\begin{equation}
\qb'_l = \Ab_l (R, \Qc(R), J_0(R), p_0(R)) \qb_l,
\label{eq:systemODE}
\end{equation}
where $\Ab_l$ is the $6 \times 6$ coefficient matrix. The explicit form of nonvanishing entries of $\Ab_l$ are reported in Appendix \ref{coeff}.

The linearized boundary conditions can be immediately derived by expanding \eqref{BCc}, \eqref{BCt} to order $O(\epsilon)$, and using \eqref{S1}, \eqref{mu1}, \eqref{solu}, \eqref{eqB} and \eqref{soluh}. By defining the functions
\begin{subequations}
\begin{equation}
g(R) :=\Omega \Pc_l - \frac{\Rc T ( 2\chi  (J_0 - 1 ) - J_0  )}{J_0^3(J_0 -1)} \left[
 \Qc J_0 R^{-1}(2 \Uc_l - l (l +1) \Vc_l)+ \Qc^{-2}{\Uc_l'}\right],
\end{equation}
\begin{equation}
f_R(R) := \Pc_l + G_d J_0 \Qc^5 R^{-1}[2 \Uc_l - l (l + 1) \Vc_l]- G_d \Qc^2 \Uc_l',
\end{equation}
\begin{equation}
f_\Th(R) := -G_d(l + 1)R^{-1} [J_0 \Qc^3 (\Uc_l - \Vc_l) + R \Vc_l'],
\end{equation}
\end{subequations}
the boundary conditions can be written in the form
\begin{subequations}
\begin{equation}
g(R_d) = 0, \qquad f_R(R_d) = 0, \qquad  f_\Th(R_d) = 0,
\label{eq:4.46a}
\end{equation}
\begin{equation}
g(R_c) - \Omega f_R(R_c) = 0, \qquad  f_\Th(R_c) = 0.
\label{bci}
\end{equation}
\end{subequations}
Equations \eqref{eq:4.46a} represent the vanishing of first-order chemical potential and the first-order stress components at the external boundary. Similarly, \eqref{bci}$_2$ states the vanishing of the stress tangential component on the inner boundary. By contrast, a more careful analysis is required for \eqref{bci}$_1$ which is derived from \eqref{BCc}$_2$. Actually, it comprises two separate boundary conditions, as we now discuss.

Let us compute the cavity volume perturbation due to the displacement field \eqref{solu}. By using the equation \eqref{vi} to compute the cavity volume via Nanson's formula and using \eqref{eq:uv1}, up to $O(\epsilon)$, we get
\begin{equation}
v_c = \frac{4}{3} \pi  r_c^3 + \epsilon \frac{2}{3} \pi R_c^2 \int_0^\pi [  (3 u + \pt_\Th v) \sin \Th + v \cos \Th ]  \rm d \Th\,.
 \label{eq:vol1}
\end{equation}
The substitution of \eqref{solu} into \eqref{eq:vol1}, shows that to first order the perturbation of $v_c$ vanishes, for any incremental displacement field. As a consequence  the hydrostatic pressure  of the enclosed liquid also remain unchanged up the first order and, hence, \eqref{bci}$_1$ is replaced by the two conditions
\begin{equation}
g(R_c)=0, \qquad  f_R(R_c) = 0.
\label{eq:bc2}
\end{equation}
In so doing, the system of six first-order linear equations \eqref{eq:systemODE} is complemented with the six boundary conditions \eqref{eq:4.46a}, \eqref{bci}$_2$ and \eqref{eq:bc2}.

\subsection{Critical volumes and bifurcation modes	}
We recall that the ratio between the actual and the swollen volume enclosed by the spherical shell, that is, the cavity volume, is measured by the parameter $\beta$, defined in \eqref{eq:beta}. For $\beta <1$, that is, for shrinking cavities, we expect to observe a bifurcation from the spherical solution to a new shape, since the spherical configuration becomes unstable.

Our problem is to find the values of $\beta$ for which there exist nontrivial solutions of our system of ordinary differential equations. As described in \S\ref{sec:linearized}, the coefficients of these incremental equations are determined by the solution of the zeroth-order problem. It is important to observe that the critical parameter $\beta$ appears explicitly only in the boundary conditions of the zeroth-order problem, see \eqref{vcrc}, \eqref{bc:r} and \eqref{eq:beta}. This couples the $O(1)$ and the $O(\epsilon)$ problems so that, in order to find the critical value $\beta_c$, we need to solve the zeroth-order equations \eqref{eq:constraint},\eqref{bf},\eqref{eq:h0} in the unknowns $(r, p_0, J_0)$ and the first-order equations \eqref{eq:systemODE} in the unknowns $(\Uc_{l}, \Uc'_{l}, \Vc_{l}, \Vc'_{l}, \Pc_{l}, \Hc_{l})$, simultaneously.\\
To this end, let $\yb_{l} = \{r, p_0, J_0, \Uc_{l}, \Uc'_{l}, \Vc_{l}, \Vc'_{l}, \Pc_{l}, \Hc_{l}\}$ be a new unknown variable, where $l$ stands for the order of the Legendre polynomial. The system of ODEs which put together both the zeroth and the first order equations can be written as a system of first order equations of the form
 \begin{equation}
 \yb_{l} ' = \fb(\yb_{l}, R),
\end{equation}
where $R$ is the independent variable. The first three equations correspond to the spherical problem, they are nonlinear and affect the problem at order $O(\epsilon)$. By contrast, the remaining six equations describe the linearized problem at order $O(\epsilon)$ and do not influence the zeroth-order problem. Furthermore, there are two additional unknown constants, namely the integration constant $C_0$ (see Eqn. \eqref{eq:h0}) and the critical parameter $\beta_c$.

Therefore, we need a total of eleven boundary conditions. Ten of these are given by Eqs.\eqref{bc:0}, \eqref{bc:r}, \eqref{eq:4.46a}, \eqref{bci}$_2$ and \eqref{eq:bc2}. The eleventh boundary condition is $\Uc_{l}(R_c) = \bar \Uc$, with $\bar \Uc \neq 0$, and imposes a non-trivial solution of the problem. Since we deal with an eigenvalue problem, the particular choice of $\bar \Uc$ does not affect the result of the problem \cite{Ambrosetti}. 
\begin{figure}[h]
	\centering\includegraphics[width=5in]{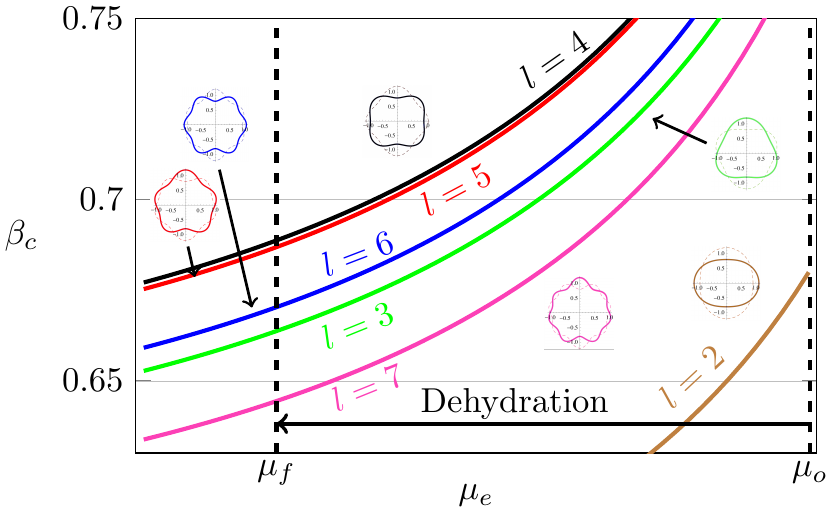}
	\caption{Critical ratio $\beta_c$  versus  $\mu_e$. Different colours indicate different modes: the order of the mode is written along the lines whereas a few insets show the shapes corresponding to the first $3$ modes. The vertical dashed lines mark the initial (right) and final (left) values of the external potential: $\mu_o=0 \,\text{J/mol}$ and  $\mu_f = -2000 \,\text{J/mol}$ which are the same used in the numerical experiments (see Fig.\ref{fig:2}).}
\label{fig:4}
\end{figure}

\begin{figure}[t]
\centerline{
\includegraphics[width=5in]{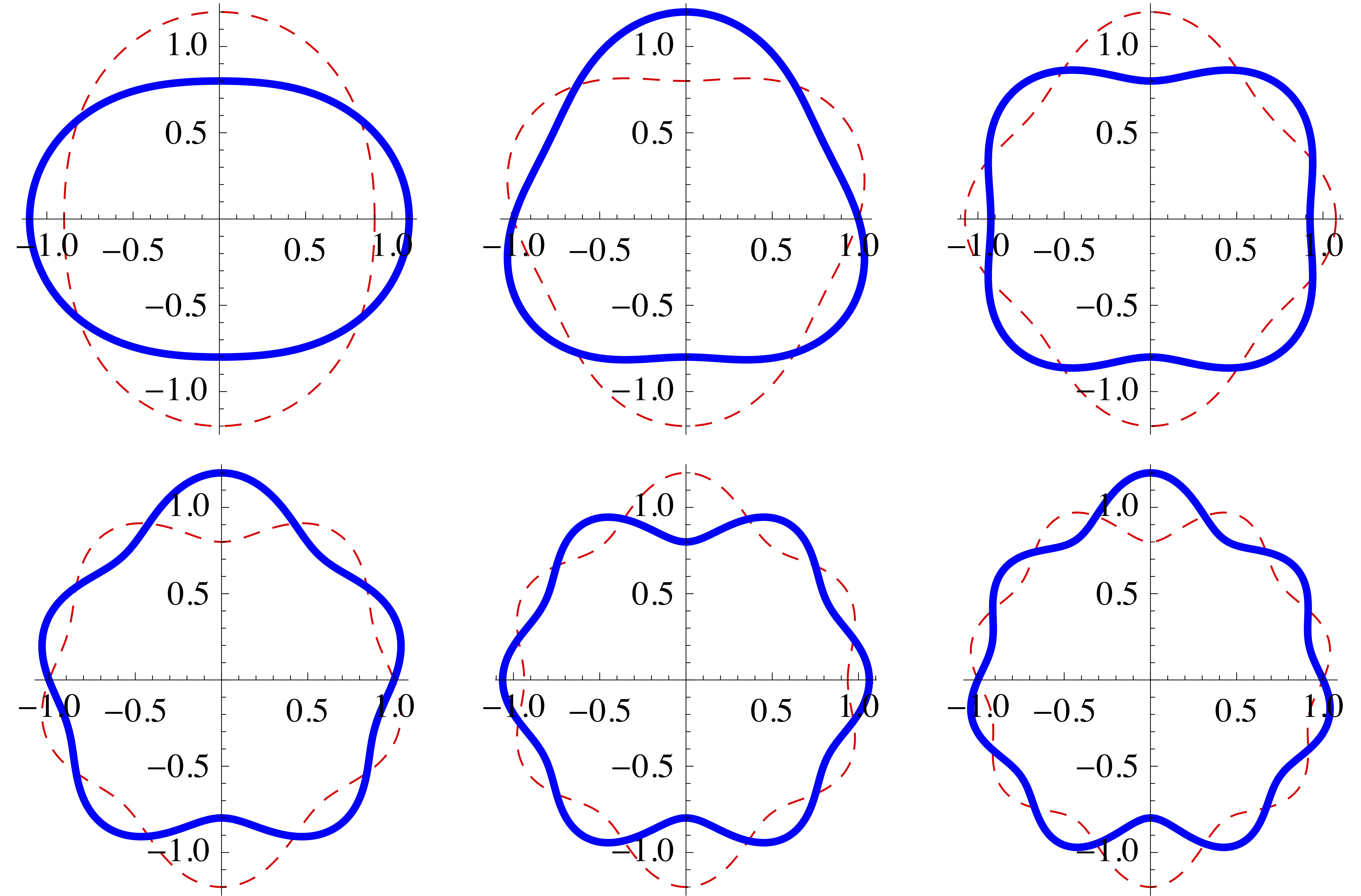}}
\caption{\label{fig:5}  Section of axisymmetric profile modes after the bifurcation. From top left to bottom right $l$ goes form 2  to 7.  The shape observed solving the complete dynamical problem in Figure \ref{fig:2} is the mode $l=6$ obtained by using the linear analysis.}
\end{figure}

\noindent
Numerical integration is performed using the {\tt Matlab} function {\tt bvp4c} which solves a boundary value problem by collocation method. Critical thresholds and corresponding shape profiles are sketched in Figure \ref{fig:5} and \ref{fig:4}, respectively. The material parameter values are taken from Table \ref{tab:1} and the size of the shell is chosen as in the numerical experiments  presented in Section \ref{S3}: $R_d = 10^{-2}\text{m}$ and $R_c/R_d = 0.875$.
Figure \ref{fig:4} shows the critical ratio $\beta_c$  versus the external chemical potential $\mu_e$ which is the control parameter of all the process. We observe that, for a given $\mu_e$, the bifurcation modes are not ordered and, for each mode $\beta_{c}$ is an increasing function of $\mu_e$. It means that whereas instability occurs for values of $\mu_e$ closer to $\mu_o$, then the cavity volume $v_c$ will be closer to the initial value $v_{co}$. Hence, what is expected as standard along a smooth de-hydration process, also holds in presence of instability. In particular,  the fourth mode is the first to be excited; however, in agreement with the classic results \cite{BenAmar:2005} which say that thinner shells develop bifurcations at higher modes, also in our case the order of modes changes according to the thickness of the capsule. It is worth noting that the intersection of the dashed line at $\mu=\mu_f$ and the blue line corresponding to mode $l=6$ yields a critical value $\beta_c = 0.67$ which is in excellent agreement with the value $0.68$ found in the dynamical simulation shown in figure \ref{fig:2} based on the numerical implementation of the full three--dimensional stress-diffusion model. Moreover, as it is shown in figure \ref{fig:5}, also the profile mode obtained by using the linear analysis coincide with the one got from full numerical analysis, up to the sign.  
Indeed, it is important to remark that the linear analysis cannot give information about neither the magnitude nor the sign of the amplitude. So, in figure \ref{fig:5} amplitudes have been chosen in order to illustrate the structure of the solution and are not related to the mechanical problem at hand. The shapes shown in solid blue line are obtained with $\bar \Uc<0$, while dashed red lines correspond to $\bar \Uc>0$. The magnitude of the solution (but not its sign) is determined by the $O(\epsilon^2)$ perturbation equations, while the analysis  of the sign requires even higher orders (see, for instance, \cite{Biscari:2004}). In the same figure \ref{fig:5}, a few 
schematic shell section profiles after the bifurcation are shown, corresponding to modes from $2$ to $7$.\\ 
%
\section{Conclusions}
We considered the instabilities of an elastic spherical shell that swells during a chemical dehydration process. The application of a difference in chemical potential induces the emptying of the cavity enclosed by the shell, providing a negative pressure on the inner wall and, at the same time, the swelling of the shell. When the enclosed volume reaches a critical value, the pressure induces an instability and the shell loses its spherical shape.

This phenomenon is captured by a finite element simulation, which solves the coupled chemo-mechanical problem. More specifically, this consists of the new Hookean elastic model for the shell and the diffusion equation for the solvent.

In order to understand how material and geometric parameters affect this instability, we performed a linear bifurcation analysis  about the spherical solution. We worked in a quasi-static regime, by assuming that the characteristic time associated with diffusion is small compared to the characteristic time of deformation. Furthermore, we also assume that the external parameter change is much slower than the diffusion time. In this approximation, the time is parameterized by the evolution of the volume enclosed by the capsule.

Our analysis was inspired by other related works that dealt with the purely mechanical instability of elastic shells under pressure \cite{Wesolowski:1967} or due to the combined effect of pressure and differential growth \cite{BenAmar:2005}. However, a key and necessary ingredient of our analysis, that makes it different from previous studies, is the introduction of the diffusion equation. In fact in our case, the mechanical and chemical problems are strongly coupled and the bifurcation is induced by the difference of the external and internal chemical potentials. Therefore, the chemo-mechanical problem is more challenging for several reasons: (i) there is a larger number of state variables; (ii) the solution with spherical symmetry cannot be determined analytically; (iii) the local volumetric deformation in the spherical solution is not uniform in space and it is a-priori unknown. We observe that a purely mechanical problem, with a neo-hookean incompressible shell with fixed enclosed volume, would lead to an overestimated critical value $\beta_c \approx 0.85$ as opposed to our $\beta_c \approx 0.67$.

Despite the richness of the model, the perturbed solution still has a classical mathematical structure in that it can be decomposed into the product of radial function (to be determined numerically) and an angular function written in terms of Legendre polynomials.

The thresholds obtained from the perturbative analysis successfully capture the instability observed in the FEM simulation. In particular, the critical threshold of mode $l=6$, which corresponds to the simulated post-buckling shape, shows and excellent agreement between perturbative ($\beta_c \approx 0.67$) and numerical ($\beta_c \approx 0.68$) values. Furthermore, the perturbation analysis reproduces correctly the dynamical concentration profile at the bifurcation as a function of shell radius (see Fig. \ref{fig:3}). 

Finally, we observe that the linear analysis predicts that modes $l=4$ and $l=5$ occur at a higher critical volume (hence, before) that mode $l=6$. However, this is not in contrast with the FEM simulation results where we observe mode $l=6$. In fact, since we deal with a dynamical problem, the shape evolution depends crucially on the choice of initial conditions.

\label{Dec}
%
%
\section*{Acknowledgments}
The authors would like to thank MIUR (Italian Minister for Education, Research, and University) 
and the \emph{PRIN 2017, Mathematics of active materials: From mechanobiology to smart devices},
project n. 2017KL4EF3, for financial support.


%
\section*{Appendix: Inhomogeneous capsules}
\label{Inhom}
\begin{figure}[h]
\centering\includegraphics[width=2.5in]{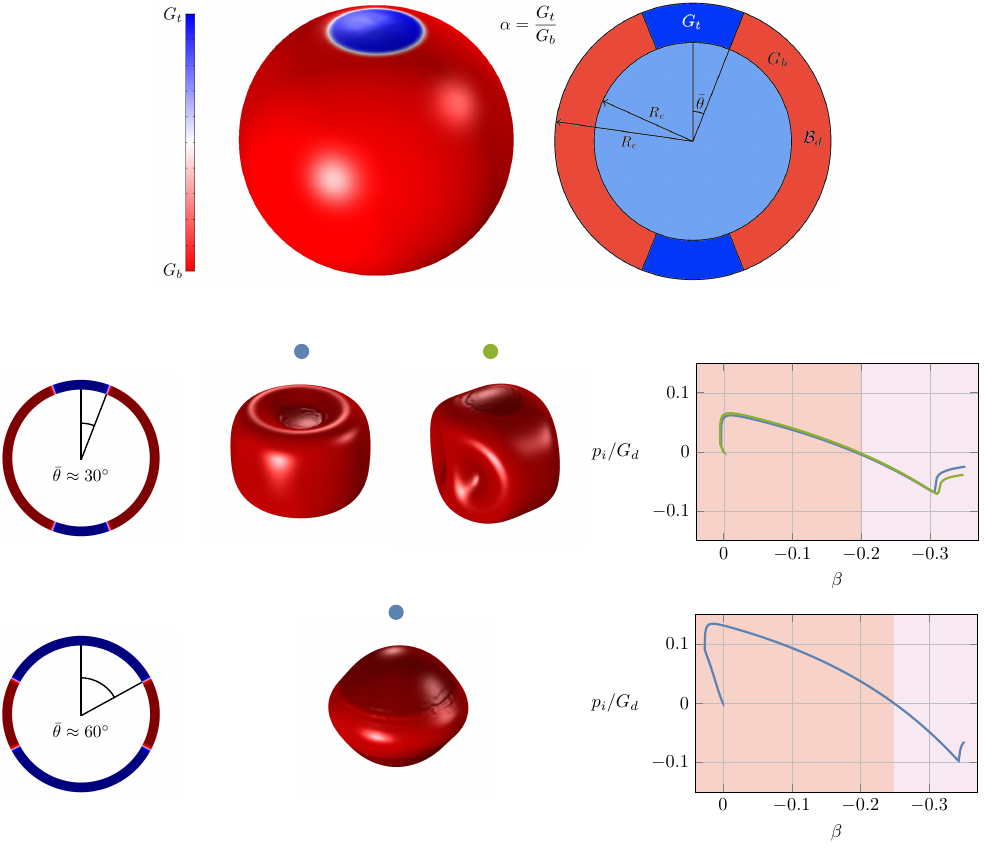}
\caption{Active-elastic instability including inhomogeneity in the material.}
\label{fig:6}
\end{figure}

A first study of mechanical instabilities in presence of inhomogeneous shells can be numerically implemented. We assume, to keep the problem simple, that the inhomogeneity of the spherical capsule can be described by two parameters: the deepness parameter $\bar{\Th}$ identifies the amplitude of two stiffer regions around the poles whereas the stiffness amplifier parameter $\alpha=G_p/G_d$ identifies the amount of increased stiffness at the poles (see Figure \ref{fig:6}, left).\\
As expected, we found that, under these conditions, the range of instability mechanisms driven by the de-hydration process is wider.  More specifically, using $\alpha=G_t/G_b=3$, we observed two de-hydration dynamics represented by the blue and orange curve in figure \ref{fig:6} (right), respectively for $\bar{\Th}=22^\circ$ and $\bar{\Th}=64^\circ$. The instability occurs at different critical pressures but also at slight different values of the ratio $v_c/v_{co}$ and the two buckled shapes are considerably different for the two cases. It is also worth noting that the shape before instability occurs is not a perfect sphere even if can be kept very close to it. The first results indicate that tuning geometry and material of the shell there is a vast range of achievable results in terms of shapes which might be controlled by $\bar{\Th}$ and $\alpha$. The study of all the achievable shapes with local inhomogeneities is far beyond the purpose of the paper but we think it is a very interesting topic and possibly a subject of our future research.
\section*{Appendix: Coefficients of the ODEs system}
\label{coeff}
We define 
\[
\gi := \frac{{\rm d} \mu_0}{{\rm d} J_0}
\]
the nonvanishng coefficients of the of the linear system to solve are
\[A_{12} = 1
\]
\[A_{21} = \frac{G_d \,\Omega  \Qc^3 \{(J_0-1)[(2 + l + l^2)  \Qc + 2 \Rc \, T\, \gi  J_0^2 \Qc^3] - 2 R[ (J_0-1)(\Rc T (J_0 \gi)' - \Omega p'_0)+ J_0 \mu'_0]\}}{ (J_0-1) (\Rc \, T\, \gi + G_d \,\Omega  \Qc^4)R^2}
     \]
  \[ A_{22} = \frac{-(J_0-1)[\Rc T J_0 (-2 \gi + 4 \gi \Qc^3 + \gip R)  + 2 \Omega \, G_d J_0 \Qc^4 - \mu'_0 R] - \mu'_0 R}
  {J_0 (J_0-1) (\Rc\; T \gi+ G_d \Omega \Qc^4) R}
  \]
  \[ A_{23} = \frac{-l (l+1) \Qc^3 \{(J_0 - 1)  [\Rc\, T (J^2_0 \Qc^3 \gi - (J_0 \gi )' R )+ 2  \Omega (G_d \Qc +  p'_0 R ) ]- J_0 \mu'_0 R\}}
  { (J_0-1) (\Rc \, T\, \gi + G_d \,\Omega  \Qc^4)R^2}
     \]
   \[A_{24} =\frac{l (l + 1) \Rc\, T\, \gi J_0 \Qc^3}{R (\Rc\, T\, \gi + G_d \Omega \Qc^4)} , \qquad
   A_{26} =  -\frac{\Rc\, T\, \Omega  J_0^2 \Qc^6}{
 D ( J_0-1 ) (\Rc\, T\, \gi + G_d\Omega \Qc^4)}, 
  \]
   \[
   A_{34} = 1,
   \]
   \[
   A_{41} = - \frac{2}{R^2} - \frac{p'_0}{G_d  \Qc R}, \qquad A_{43} = \frac{l (l + 1)}{R^2} + \frac{p'_0}{G_d  \Qc R},
   \]
   \[
   A_{44} = - \frac{2}{R}, \qquad A_{45} = \frac{\Qc J_0}{G_d R},
   \]
   \begin{align*}
   A_{51} & = \big( (J_0-1) (\Rc\, T\, \gi+ G_d \Omega \Qc^4) R^2\big)^{-1}
   \{\Rc\, T\, \gi \Qc (J_0-1) \times \\
    & [-G_d\Qc (2 + l+ l^2 - 2 J_0^2 \Qc^6 + 2 R \Qc^3 J'_0) - 2 p'_0 R] - 2 R G_d J_0 \Qc^5 [\Rc\, T\, ( J_0-1) \gip+ \mu'_0]\},
   \end{align*}
   \[
   A_{52} = \frac{G_d \, \Qc^2 [-\Rc\, T\, J_0 ( J_0- 1) (4 \gi (-1 + J_0 \Qc^3) +  \gip R) + 
   (-2 + J_0) \mu'_0 R ]}{J_0  ( J_0 -1 ) (\Rc\, T\, \gi + G_d \,\Omega \Qc^4)R},
   \]
   \begin{align*}
   A_{53} & = \big((J_0 -1) (\Rc\, T\, \gi + G_d \,\Omega \Qc^4)R^2 \big)^{-1}
   \{l (l + 1) \Qc [\Rc\, T\, \gi (J_0 -1) \times \\
   & (G_d \, \Qc (2 - J_0^2 \Qc^6 +  \Qc^3 J'_0 R) + p'_0 R) + R G_d \, J_0 \Qc^4 (\Rc\, T\, ( J_0-1) \gip + \mu'_0)]\},
   \end{align*}
   \[
   A_{54} = \frac{ l (l + 1) \Rc\, T\, G_d \, \gi J_0 \Qc^5}{ (\Rc\, T\, \gi + G_d \,\Omega \Qc^4)R},
   \qquad
   A_{56} = -\frac{\Rc\, T\, G_d \,\Omega J_0^2 \Qc^8}{D ( J_0 - 1 ) (\Rc\, T\, \gi + G_d \,\Omega \Qc^4)},
   \]
  \[
   A_{61} = \frac{D \, l (l +1) (J_0 - 1) (-2 \Rc\, T \gi J_0^2 \Qc^3 + \mu'_0 R)}{ \Omega J_0 \Rc \, T  R^3},
   \qquad
   A_{62} =  -\frac{D\,\gi \, l (l + 1) (J_0-1)}{ \Omega\, R^2 },
   \]
   \[
   A_{63} = \frac{D \, l (l + 1) (J0-1) [l(l + 1) \Rc \,T\, \gi J_0^2 \Qc^3 - \mu'_0 R]}{\Omega  \Rc \, T  J_0 R^3},
   \]
   \[
   A_{64} =  \frac{D \, l (l + 1) (J_0 - 1) \mu'_0}{\Omega \Rc \, T J_0^2 \Qc^3 R},
   \qquad
   A_{65} = - \frac{D \, l (l + 1) (J_0 - 1)\Qc^2}{ \Rc \, T   R^2}
   \]

\end{document}